# Forecasting and evaluating intervention of Covid-19 in the World

Zixin Hu, Ph.D[1,2]; Qiyang Ge, MS[3] ; Shudi Li,MS[4] ;Eric Boerwincle[4] ,Ph.D[4] , Li Jin , Ph.D[1,2] and Momiao Xiong, Ph.D[4,*]

[1] State Key Laboratory of Genetic Engineering and Innovation Center of Genetics and Development, School of Life Sciences, Fudan University, Shanghai, China.

[2] Human Phenome Institute, Fudan University, Shanghai, China.

[3] The School of Mathematic Sciences, Fudan University, Shanghai, China.

[4] School of Public Health, The University of Texas Health Science Center at Houston, Houston, TX 77030, USA.

**Running Title**: Real Time Forecasting of Covid-19 in the world

**Keywords:** Cov-19, artificial intelligence, transmission dynamics, forecasting, time series, auto-encoder.

*Address for correspondence and reprints: Dr. Momiao Xiong, Department of Biostatistics and Data Science, School of Public Health, The University of Texas Health Science Center at Houston, P.O. Box 20186, Houston, Texas 77225, (Phone): 713-500-9894, (Fax): 713-500-0900, E-mail: Momiao.Xiong@uth.tmc.edu.

**Key Points**

**Question** As the Covid-19 pandemic surges around the world, what are the number of cases worldwide at its peak, the length of the pandemic before receding, the fatality rate and the timing of public health interventions to significantly stop the spread of Covid-19?

**Findings** We estimated the maximum number of cumulative cases under immediate intervention to be 1,530,276; under later intervention the number increased to a frightening 255,392,154 and the case ending time was delayed to January 10, 2021.

**Meaning** If there is not immediate aggressive action to intervene, we will face serious consequences.

**Abstract**

**Importance** When the Covid-19 pandemic enters dangerous new phase, whether and when to take aggressive public health interventions to slow down the spread of COVID-19.

**Objective** To develop the artificial intelligence (AI) inspired methods for real-time forecasting and evaluating intervention strategies to curb the spread of Covid-19 in the World.

**Design, Setting, and Participants** A modified auto-encoder for modeling the transmission dynamics of the epidemics is developed and applied to the surveillance data of cumulative and new Covid-19 cases and deaths from WHO, as of March 16, 2020.

**Interventions and Exposures** Pubic health interventions and documented novel coronavirus (2019-nCoV)–infected pneumonia (NCIP).

**Main Outcomes and Measures** Data on the number of lab confirmed cumulative, new and death cases of Covid-19 from January 20, 2020 to March 16 were obtained from WHO.

**Results** The average errors of 5-step forecasting were 2.5%. The total peak number of cumulative cases and new cases, and the maximum number of cumulative cases in the world with later intervention (comprehensive public health intervention is implemented 4 weeks later) could reach 75,249,909, 10,086,085, and 255,392,154, respectively. The case ending time was January 10, 2021. However, the total peak number of cumulative cases and new cases and the maximum number of cumulative cases in the world with one week later intervention were reduced to 951,799, 108,853 and 1,530,276, respectively. Duration time of the Covid-19 spread would be reduced from 356 days to 232 days. The case ending time was September 8, 2020. We observed that delaying intervention for one month caused the maximum number of cumulative cases to increase 166.89 times, and the number of deaths increase from 53,560 to 8,938,725.

**Conclusions and Relevance**    Complimentary to other approaches to model the transmission dynamics of the virus outbreaks, the AI-based method provides forecasting tools for tracking, estimating the trajectory of epidemics, assessing their severity, and predicting the lengths of epidemics with various interventions. We estimated the maximum number of cumulative cases under immediate intervention to be 1,530,276; under later intervention the number increased to a frightening 255,392,154 and the case ending time was delayed to January 10, 2021.

**Introduction**

As of March 16, 2020, global confirmed cases of COVID-19 passed 170,568 and has spread to more than 152 countries. As this coronavirus passes into a pandemic,[1] a number of questions arise among the public and government and business leaders: How many cases will there be worldwide? How many deaths can we expect? When will we see a peak in the number of cases? When will this pandemic end? And, how will recommended immediate actions slow the spread?

A number of the statistical and dynamic models of the Covid-19 outbreak have been applied to analyze its transmission dynamics.[2-8] These epidemiological models are useful for estimating the dynamics of transmission, targeting resources and evaluating the impact of intervention strategies, but the models require values for unknown parameters and depend on many assumptions,[8-10] leading to low accuracy and unsure prediction.

**Methods**

**Modified Auto-encoder for Modeling Time Series**

To overcome these limitations, we developed a modified auto-encoder (MAE) (eFigure 1),[11] an AI based method for real time forecasting of the new and cumulative confirmed cases of Covid-19 under various interventions around the world.[12-13] Transfer learning was used to train the MAE.[13] Weights between 0 and 1 were assigned to an intervention variable for the different degrees of interventions – zero being no intervention and one being complete. Complete intervention included keeping social distance, washing hands, strict travel restriction, no large group gatherings, mandatory quarantine, restricted public transportation, keep social distance, school closing and closure of all non-essential business, including manufacturing. Using real data to evaluate the consequences of specific intervention is infeasible. We considered four intervention scenarios. The first intervention scenario started intervention on March 24, 2020 with a weight of 0.5, one week later transitioned to a complete intervention with a weight of 1. The second intervention started with a weight of 0 for the first week, a weight of 0.5 for the second week, and two weeks later transitioned to a complete intervention with a weight of 1. The third intervention was delaying two weeks, limited in the thrid week and complete in the fourth week. The fourth intervention delayed actions for three weeks, limited in the fourth weeks and complete in the fifth week. For each scenario, we investigated how the degree and timing of the intervention determined the peak time and case ending time, the peak number and maximum number of cases and forecasting the peak and maximum number of new and cumulative cases in more than 150 countries across the world.

**Data Collection**

The analysis is based on the surveillance data of confirmed cumulative and new Covid-19 cases in the world up to March 16, 2020. Data on the number of cumulative, new and death cases of Covid-19 from January 20, 2020 to March 16 were obtained from WHO (https://www.who.int/emergencies/diseases/novel-coronavirus-2019/situation-reports). Data included the total numbers of cumulated, newly confirmed and attributed deaths across 152 countries.

**Results**

**Later intervention makes it difficult to stop the spread of Covid-19**

To demonstrate that the MAE is an accurate forecasting method, the MAE was applied to confirmed accumulated cases of Covid-19 across 152 countries. The intervention indicator for

China and other countries was set to 1 and 0, respectively. eTable 1 presented the one-step to five-step errors for forecasting the cumulative cases, starting from March 12, 2020. In all cases, the average forecasting accuracies of the MAE were less than 2.5% (eTable 1).

Table 1 shows the forecasting results of Covid-19 in 30 countries and worldwide under later intervention (Scenario 4). The total peak number of cumulative cases and new cases in the world with later intervention could reach 75,249,909 and 10,086,085, respectively. If every country in the world undertook later intervention, the total number of cases in the world could reach as high as 255,392,154 and community transmission of Covid-19 continuing until January 10, 2021. The top 10 countries with a high average increasing number of cases were Italy, Spain, Iran, Germany, USA, France, Switzerland, Belgium, UK and Austria. To show the dynamics of COVID-19 development, eFigure 2 (G) and (H) plot curves of the number of cumulative cases and new cases in seven major infected countries: Iran, Spain, Italy, Germany, USA, France and China under scenario 4.

**New strategies are needed to curb the spread of COVID-19**

There is urgent need to develop new strategies to curb the spread of COVID-19.[1] We investigate whether complete interventions can control the spread of COVID-19 and how early complete interventions will reduce the peak time and the number, and the final total number of cases across the world. Table 2 shows the forecasted results of COVID-19 in 30 countries and worldwide under early complete intervention (Scenario 1). We can observe dramatic reduction of the cases of COVID-19. The forecasted total number of cases in the world was reduced by early complete intervention to 1,530,276 from nearly 255 million based on 4 weeks delay intervention, In other words, 99.4% of the potential cases would be eliminated by early complete intervention. The duration time was reduced from 356 days to 232 days, and the end time would change from January 10, 2021 to September 8, 2020. eFigure 2 (A) and (B) plot curves of the number of cumulative cases and new cases in six major infected countries: Iran, Spain, Italy, Germany, USA and France under scenario 1.

To investigate intervention measures between early complete and 4 week delay intervention, eTables 2 and 3 show the results under scenarios 2 and 3, respectively. eFigure 2(C), (D), (E) and (F) plot transmission dynamics of Covid-19 with plotted curves of the cumulative cases and new cases in the six major infected countries under scenarios 2 and 3, respectively.

**Comparisons among intervention strategies**

To further illustrate the impact of interventions on the spread of COVID-19, we compared the effects of the four intervention scenarios on the transmission dynamics of Covid-19 across the world. Figure 1 plots the world reported and forecasted time curves of the cumulative and newly confirmed cases of Covid-19 under the four intervention scenarios. The ratios of the world number of final cases across the four scenarios were 1:4.26:19.16:166.9 and ratios of case duration under the four intervention scenarios were 1:1:01.1.38:1.53. These results demonstrate that intervention time delays have serious consequences. 4 week delay intervention may increase the number of cases by 166.9 fold and will delay the ending time by almost four months. Delaying intervention will substantially speed the spread of coronavirus.

e Figure 3 plots the time-case curves for the top six infected countries: Iran, Spain, Italy, Germany, USA, France and China. The time-case curve under the 4 week delay intervention was shifted more than one month to the right and was much steeper than that of under the early intervention. Delaying intervention will substantially increase the number of cumulative cases of Covid-19.

eFigure 4 shows the case-fatality rate curve as a function of the Date where the case-fatality rate was defined as the ratio of the number of deaths over the number of cumulative cases in the world. The average case-fatality rate was 3.5%.

**Discussion**

Real-time forecasting are more accurate than epidemiologic transmission model where the model parameters may not applicable in practice. We estimated the duration, peak time and ending time, peak number and maximum number of cumulative cases of Covid-19 under four intervention scenarios for 152 countries in the world. This provided critical information for government and health authorities to consider urgent public health response planning to slow the spread of Covid-19. We demonstrated that aggressive interventions are urgently needed. Otherwise, we will face disastrous consequences.

**References**

1. Callaway E, Time to use the p-word? Coronavirus enters dangerous new phase, Nature (2020) doi: 10.1038/d41586-020-00551-1
2. Li Q, Guan X, Wu P, et al. Early transmission dynamics in Wuhan, China, of novel Coronavirus-infected pneumonia. N Engl J Med. Jan 29. doi: 10.1056/NEJMoa2001316. [Epub ahead of print].
3. Wu JT, Leung K, Leung GM. Nowcasting and forecasting the potential domestic and international spread of the 2019-nCoV outbreak originating in Wuhan, China: a modelling study. Lancet. 2020; 395(10225):689-697.
4. Zhao S, Musa SS, Lin Q et al. Estimating the unreported number of novel Coronavirus (2019-nCoV) cases in China in the first half of January 2020: A data-driven modelling analysis of the early outbreak. J Clin Med. 2020 Feb 1;9(2). pii: E388. doi: 10.3390/jcm9020388.
5. Kucharski A, Russell T, Diamond C, Liu Y, CMMID nCoV working group, Edmunds J, Funk S, Eggo R. Analysis and projections of transmission dynamics of nCoV in Wuhan (2020) https://cmmid.github.io/ncov/wuhan_early_dynamics/index.html.
6. Tuite AR, Fisman DN. Reporting, epidemic growth, and reproduction numbers for the 2019 novel coronavirus (2019-nCoV) epidemic. Ann Intern Med. 2020 Feb 5. doi: 10.7326/M20-0358. [Epub ahead of print].
7. Hellewell J, Abbott S, Gimma A, et al. Centre for the Mathematical Modelling of Infectious Diseases COVID-19 Working Group, Funk S1, Eggo RM2. Feasibility of controlling COVID-19 outbreaks by isolation of cases and contacts. Lancet Glob Health. 2020 Feb 28. pii: S2214-109X(20)30074-7. doi: 10.1016/S2214-109X(20)30074-7. [Epub ahead of print].
8. Li R, Pei S, Chen B, Song Y, Zhang T, Yang W, Shaman J. Substantial undocumented infection facilitates the rapid dissemination of novel coronavirus (SARS-CoV2). Science. 2020 Mar 16. pii: eabb3221. doi: 10.1126/science.abb3221. [Epub ahead of print].
9. Funk S, Camacho A, Kucharski AJ, Eggo RM, Edmunds WJ. Real-time forecasting of infectious disease dynamics with a stochastic semi-mechanistic model. Epidemics. 2018; 22:56-61.

10. Johansson MA, Apfeldorf KM, Dobson S et al. An open challenge to advance probabilistic forecasting for dengue epidemics. Proc Natl Acad Sci U S A. 2019;116(48):24268-24274.

11. Charte D, Charte F, García S, Jesus MJD, Herrera F. A practical tutorial on autoencoders for nonlinear feature fusion: Taxonomy, models, software and guidelines. Information Fusion. 2019; 44:78–96.

12. Yuan X, Huang B, Wang Y, Yang C, Gui W. Deep Learning-based feature representation and its application for soft censor modeling with variable-wise weighted SAE. IEEE Trans on Industrial informatics. 2018; 14(7): 3235-3243.

13. F. Zhuang Z. Qi K. Duan D et al. He. A Comprehensive Survey on Transfer Learning. arXiv:1911.02685. (2019)

**Legend**

**Figure 1.** The reported and forecasted curves of the cumulative and new confirmed cases of Covid-19 in the world as a function of days from January 20, 2020 to July 28, 2020.

Table 1. Spread of Covid-19 in 30 countries and world wide under 4 weeks delay intervention.

| State | Peak Time | End Time | Duration | Peak (Cum) | Peak (New) | Current Case | End Case |
|---|---|---|---|---|---|---|---|
| Total | 4/17/2020 | 1/10/2021 | 356 | 75249909 | 10086085 | 170568 | 255392154 |
| ITALY | 4/17/2020 | 1/10/2021 | 346 | 14945480 | 1999429 | 24747 | 53281848 |
| SPAIN | 4/17/2020 | 1/10/2021 | 345 | 10080564 | 1351788 | 7753 | 33196999 |
| IRAN | 4/17/2020 | 1/6/2021 | 322 | 8556153 | 1146663 | 14991 | 27343905 |
| GERMANY | 4/17/2020 | 1/10/2021 | 349 | 6532219 | 875856 | 4838 | 21864400 |
| USA | 4/17/2020 | 1/10/2021 | 356 | 4532725 | 607493 | 4740 | 16644849 |
| FRANCE | 4/17/2020 | 1/10/2021 | 352 | 4263429 | 572051 | 5380 | 14555999 |
| SWIZTERLAND | 4/17/2020 | 1/10/2021 | 320 | 3092785 | 414952 | 2200 | 9772913 |
| BELGIUM | 4/17/2020 | 1/5/2021 | 336 | 2835657 | 380783 | 1085 | 8727195 |
| UK | 4/17/2020 | 1/10/2021 | 345 | 1624266 | 218542 | 1395 | 6349494 |
| AUSTRIA | 4/17/2020 | 1/10/2021 | 320 | 1156505 | 156173 | 959 | 4206694 |
| NORWAY | 4/17/2020 | 1/10/2021 | 319 | 1214800 | 163068 | 1077 | 3894919 |
| MALAYSIA | 4/17/2020 | 1/10/2021 | 351 | 1081414 | 144904 | 553 | 3750555 |
| GREECE | 4/17/2020 | 11/4/2020 | 252 | 1047665 | 141301 | 331 | 3595859 |
| NETHERLANDS | 4/17/2020 | 1/10/2021 | 318 | 881147 | 118402 | 1135 | 3080802 |
| PORTUGAL | 4/17/2020 | 1/10/2021 | 314 | 675964 | 91093 | 245 | 2104149 |
| FINLAND | 4/17/2020 | 1/10/2021 | 347 | 578886 | 77668 | 267 | 1923049 |
| ESTONIA | 4/17/2020 | 1/10/2021 | 319 | 607872 | 81796 | 205 | 1902652 |
| SLOVENIA | 4/17/2020 | 1/10/2021 | 312 | 598294 | 80475 | 219 | 1891314 |
| ISRAEL | 4/17/2020 | 1/10/2021 | 324 | 526864 | 71296 | 200 | 1867519 |
| CANADA | 4/17/2020 | 1/10/2021 | 350 | 480352 | 64450 | 304 | 1792760 |
| CZECHIA | 4/17/2020 | 1/8/2021 | 313 | 500323 | 67284 | 298 | 1708210 |
| ICELAND | 4/17/2020 | 1/10/2021 | 315 | 438161 | 59381 | 138 | 1570527 |
| ROMANIA | 4/17/2020 | 1/10/2021 | 319 | 383176 | 51910 | 158 | 1389549 |
| QATAR | 4/17/2020 | 1/10/2021 | 316 | 428531 | 57690 | 401 | 1245999 |
| BRAZIL | 4/17/2020 | 1/6/2021 | 315 | 374378 | 50246 | 200 | 1218993 |
| AUSTRALIA | 4/17/2020 | 1/10/2021 | 352 | 353747 | 47491 | 298 | 1190874 |

| KOREA  | 4/17/2020 | 10/30/2020 | 284 | 296036 | 38849 | 8236 | 1019408 |
| POLAND | 4/17/2020 | 1/5/2021   | 308 | 287008 | 38725 | 150  | 985182  |

Table 2. Spread of Covid-19 in 30 countries and world wide under early complete intervention (one week from March 16 intervention).

| State | Peak Time | End Time | Duration | Peak (Cum) | Peak (New) | Current Case | End Case |
|---|---|---|---|---|---|---|---|
| Total | 2020/3/28 | 2020/9/8 | 232 | 951799 | 108853 | 170568 | 1530276 |
| ITALY | 2020/3/27 | 2020/9/8 | 222 | 161276 | 19998 | 24747 | 261790 |
| SPAIN | 2020/3/28 | 2020/8/20 | 202 | 117400 | 15268 | 7753 | 187157 |
| IRAN | 2020/3/27 | 2020/6/14 | 116 | 95104 | 12039 | 14991 | 157269 |
| GERMANY | 2020/3/28 | 2020/7/22 | 177 | 73998 | 9933 | 4838 | 129654 |
| USA | 2020/3/27 | 2020/6/6 | 138 | 47058 | 6454 | 4740 | 83921 |
| CHINA | 2020/2/5 | 2020/4/29 | 100 | 31432 | 5236 | 81077 | 83103 |
| FRANCE | 2020/3/27 | 2020/8/2 | 191 | 45186 | 5933 | 5380 | 81593 |
| SWITZERLAND | 2020/3/28 | 2020/9/6 | 194 | 34665 | 5031 | 2200 | 61734 |
| BELGIUM | 2020/3/28 | 2020/6/4 | 121 | 29479 | 4487 | 1085 | 52925 |
| UK | 2020/3/28 | 2020/6/2 | 123 | 19348 | 2467 | 1395 | 31006 |
| NORWAY | 2020/3/28 | 2020/6/22 | 117 | 13631 | 1985 | 1077 | 26386 |
| AUSTRIA | 2020/3/29 | 2020/6/5 | 101 | 14394 | 1825 | 959 | 24550 |
| GREECE | 2020/3/29 | 2020/6/10 | 105 | 13525 | 1922 | 331 | 22467 |
| MALAYSIA | 2020/3/28 | 2020/7/4 | 161 | 11271 | 1705 | 553 | 20985 |
| NETHERLANDS | 2020/3/26 | 2020/5/13 | 76 | 8097 | 1232 | 1135 | 16080 |
| KOREA | 2020/2/29 | 2020/5/22 | 123 | 3150 | 813 | 8236 | 15649 |
| PORTUGAL | 2020/3/30 | 2020/6/2 | 92 | 8578 | 1157 | 245 | 14841 |
| FINLAND | 2020/3/30 | 2020/7/22 | 175 | 7707 | 1037 | 267 | 13817 |
| ESTONIA | 2020/3/30 | 2020/6/21 | 116 | 7928 | 1036 | 205 | 13382 |
| SLOVENIA | 2020/3/28 | 2020/6/28 | 116 | 5856 | 958 | 219 | 12717 |
| ISRAEL | 2020/3/29 | 2020/6/30 | 130 | 5865 | 878 | 200 | 10838 |
| ICELAND | 2020/3/29 | 2020/6/23 | 114 | 4854 | 800 | 138 | 10679 |
| CZECHIA | 2020/3/28 | 2020/6/20 | 111 | 5653 | 772 | 298 | 9586 |
| CANADA | 2020/3/28 | 2020/7/8 | 164 | 5330 | 739 | 304 | 9282 |

| | | | | | | |
|---|---|---|---|---|---|---|
| QATAR | 2020/3/29 | 2020/6/11 | 103 | 4794 | 652 | 401 | 8206 |
| ROMANIA | 2020/3/29 | 2020/5/21 | 85 | 4158 | 627 | 158 | 7754 |
| AUSTRALIA | 2020/3/28 | 2020/6/13 | 141 | 4117 | 585 | 298 | 7430 |
| BRAZIL | 2020/3/28 | 2020/6/11 | 106 | 4017 | 584 | 200 | 7162 |
| DENMARK | 2020/3/12 | 2020/6/19 | 114 | 615 | 353 | 898 | 6083 |

## Materials and Methods

### Data Sources

Data on the confirmed, new and death cases of Covid-19 from January 20, 2020 to March 16 were from WHO (https://www.who.int/emergencies/diseases/novel-coronavirus-2019/situation-reports). Data included the total numbers of accumulated, new confirmed and death cases in the world and across 152 countries. The data were organized in a matrix with the rows representing the whole world and countries and columns representing the number of the new confirmed cases each day. The confirmed cases of each country were a time series. Let $t_{ij}$ be the number of the confirmed cases of the $j^{th}$ day within the $i^{th}$ country. Let $Z$ be a $I \times m$ dimensional matrix. The element $Z_{ij}$ is the number of the confirmed new cases of Covid-19 on the $j^{th}$ day, starting with January 20, 2020 in the $i^{th}$ country.

### Modified Auto-encoder for Modeling Time Series

Modified auto-encoders (MAE) (1,2) were used to forecast the number of the accumulative and new confirmed cases of Covid-19. Unlike the classical auto-encoder where the number of nodes in the layers usually decreases from the input layer to the latent layers, the numbers of the nodes in the input, the first latent layer, the second latent layer and output layers in the MSAE were 8, 32, 4 and 1, respectively (eFigure 1). We view a segment of time series with 8 days as a sample of data and take 128 segments of time series as the training samples. One element from the data matrix $Z$ is randomly selected as a start day of the segment and select its 7 successive days as the other days to form a segment of time series. Let $i$ be the index of the segment and $j_i$ be the column index of the matrix $Z$ that was selected as the starting day. The $i^{th}$ segment time series can be represented as $\{Z_{j_i}, Z_{j_i+1}, \ldots, Z_{j_i+7}\}$. Data were normalized to $X_{j_i+k} = \frac{Z_{j_i+k}}{S}, k = 0, 1, \ldots, 7$, where $S = \frac{1}{8}\sum_{k=0}^{7} Z_{j_i+k}$. Let $Y_i = \frac{Z_{j_i+8}}{S}$ be the normalized number of cases to forecast. If $S = 0$, then set $Y_i = 0$. The loss function was defined as

$$L = \sum_{i=1}^{128} W_i (Y_i - \hat{Y}_i)^2,$$

where $Y_i$ was the observed number of the cases in the forecasting day of the $i^{th}$ segment time series and $\hat{Y}_i$ was its forecasted number of cases by the MAE, and $W_i$ were weights. If $j_i$ was in the interval [1, 12], then $W_i = 1$. If $j_i$ was in the interval [13, 24], then $W_i = 2$, etc. The back propagation algorithm was used to estimate the weights and bias in the MSAE. Repeat training processed 5 times. The average forecasting $\hat{Y}_i, i = 1, \ldots, 152$ will be taken as a final forecasted number of the accumulated confirmed cases for each country.

**Forecasting Procedures**

The trained MAE was used for forecasting the future number of cumulative cases of Covid-19 for each country. Consider the $i^{th}$ country. Assume that the number of new confirmed cases of Covid-19 on the $j^{th}$ day that needs to be forecasted is $x_{ij}$. Let $H$ be a $152 \times 8$ dimensional matrix and $h_{il} = x_{ij-9+l}, i = 1, \ldots, 152$, and $l = 1, \ldots, 8$. Let $g_i = \frac{1}{8}\sum_{l=1}^{8} h_{il}, i = 1, \ldots, 152$ be the average of the $i^{th}$ row of the matrix $H$. Let $U$ be the normalized matrix of $H$ where $u_{il} = \frac{h_{il}}{g_i}, i = 1, \ldots, 152$, and $l = 1, \ldots, 8$. The output of the MAE is the forecasted number of new confirmed cases and is denoted as $\hat{v}_i = f(u_{i1}, \ldots, u_{i8}, \theta), i = 1, \ldots, 152$, where $\theta$ represented the estimated parameters in the trained MAE. The one-step forecasting of the number of new confirmed cases of Covid-19 for each country is given by $\hat{Y}_i = \hat{v}_i g_i, i = 1, \ldots, 152$.

The recursive multiple-step forecasting involved using a one-step model multiple times where the prediction for the preceding time step was used as an input for making a prediction on the following time step. For example, for forecasting the number of new confirmed cases for the one more next day, the predicted number of new cases in one-step forecasting would be used as an observational input in order to predict day 2. Repeat the above process to obtain the two-step forecasting. The summation of the final forecasted number of new confirmed cases for each country was taken as the prediction of the total number of new confirmed cases of Covid-19 worldwide.

**eFigure 1** Architecture of modified autoencoder.

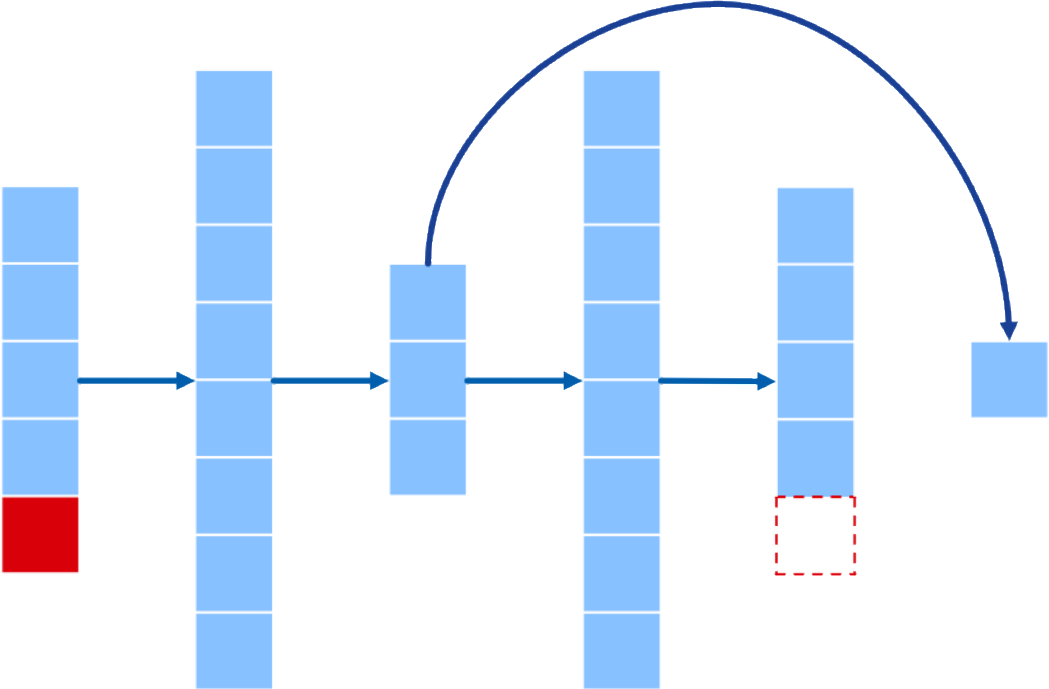

**eFigure 2** Trajectory of COVID-19 in the seven most infected countries: Iran, Spain, Italy, Germany, USA, France and China as a function of days from January 21 to June 19, 2020. (A), (C), (E) and (G) Forecasted curves of the newly confirmed cases of COVID-19 under scenarios 1, 2, 3 and 4, respectively. (B), (D), (F) and (H) Forecasted curves of the cumulative confirmed cases of COVID-19 under scenarios 1, 2, 3 and 4, respectively.

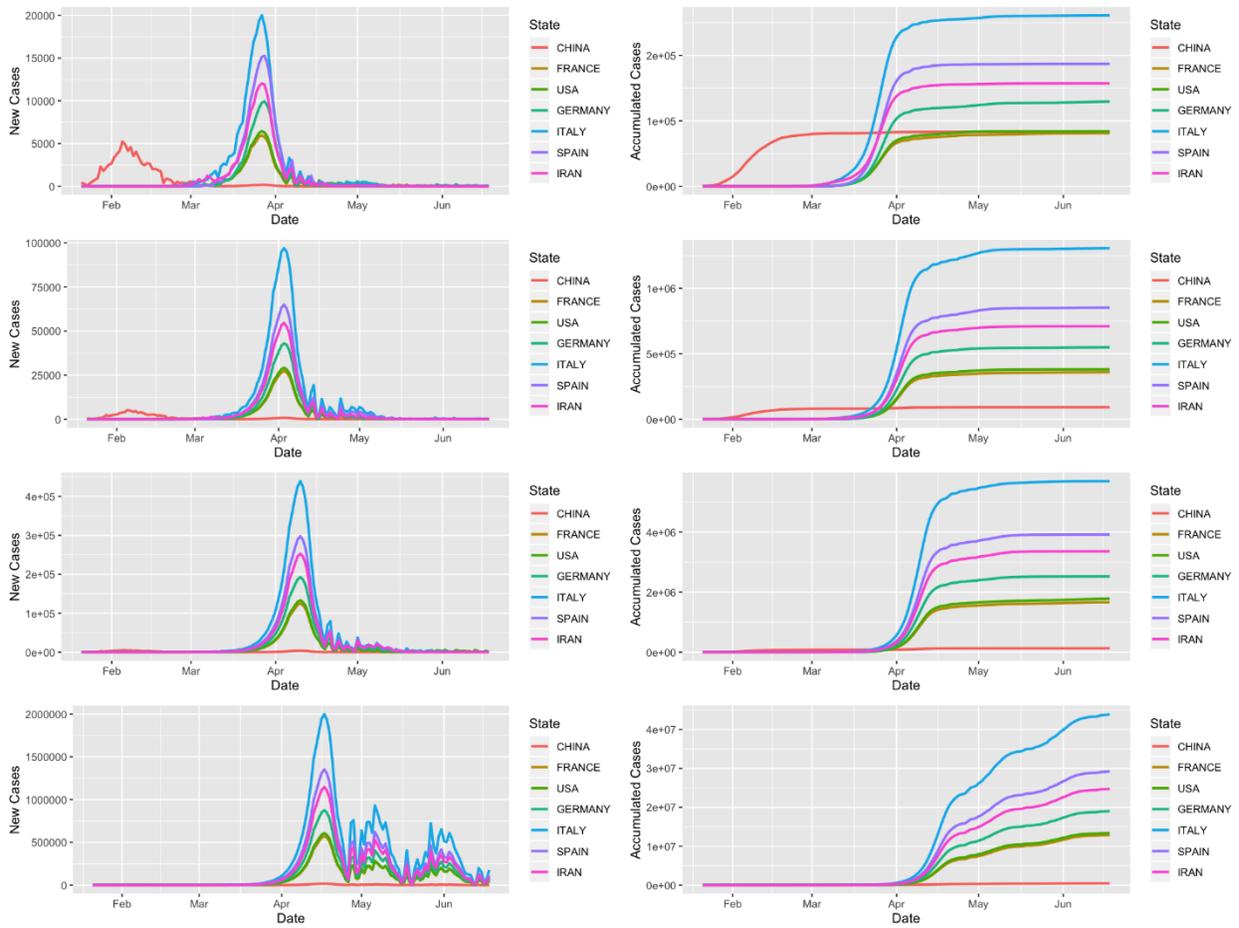

**eFigure 3** Time-case plot of the top seven infected countries: Iran, Spain, Italy, Germany, USA, France and China. (A) Time-case plot under intervention scenario 1; (B) Time-case plot under intervention scenario 2; (C) Time-case plot under intervention scenario 3 and (D) Time-case plot under intervention scenario 4.

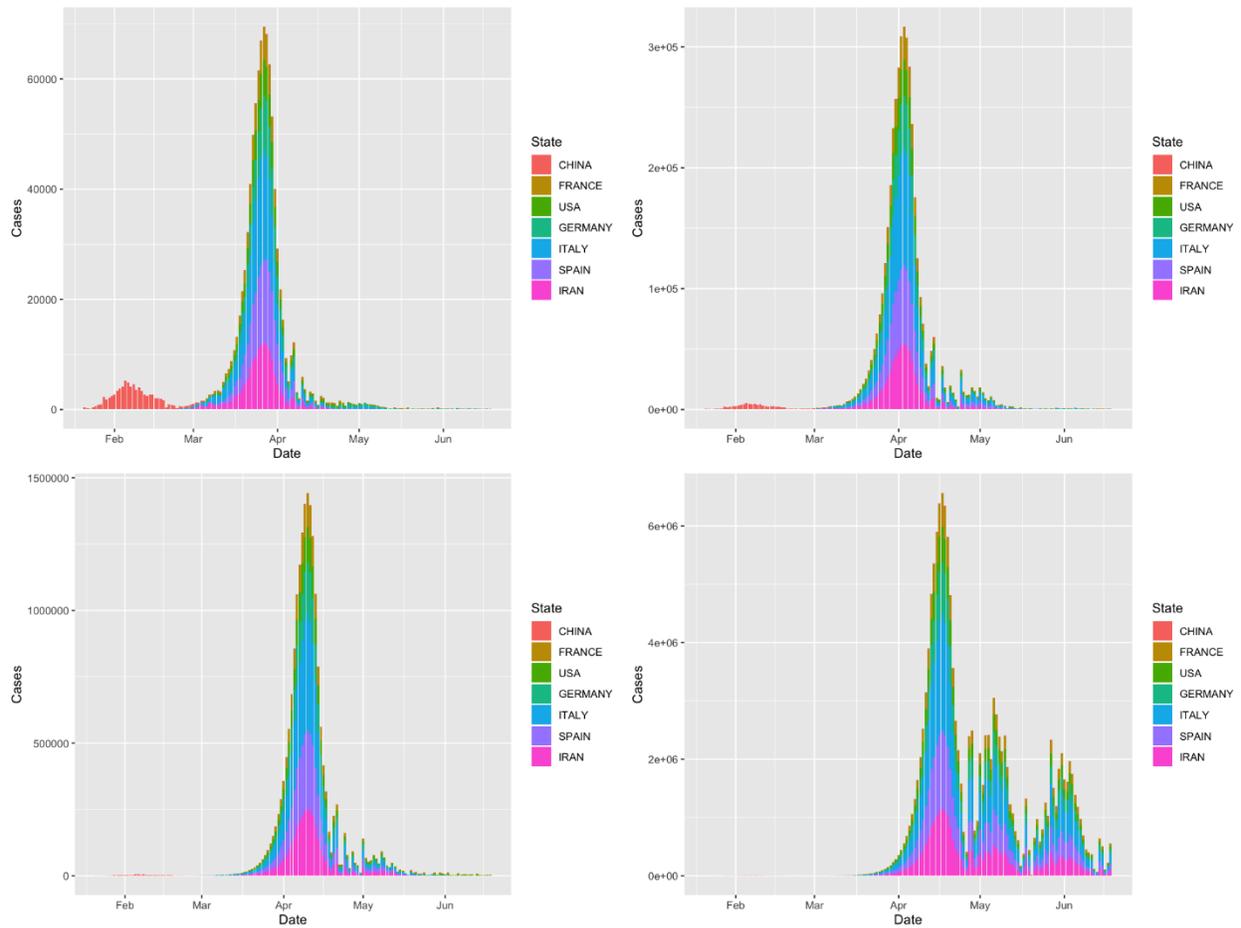

**eFigure 4** Case-fatality curve for the world.

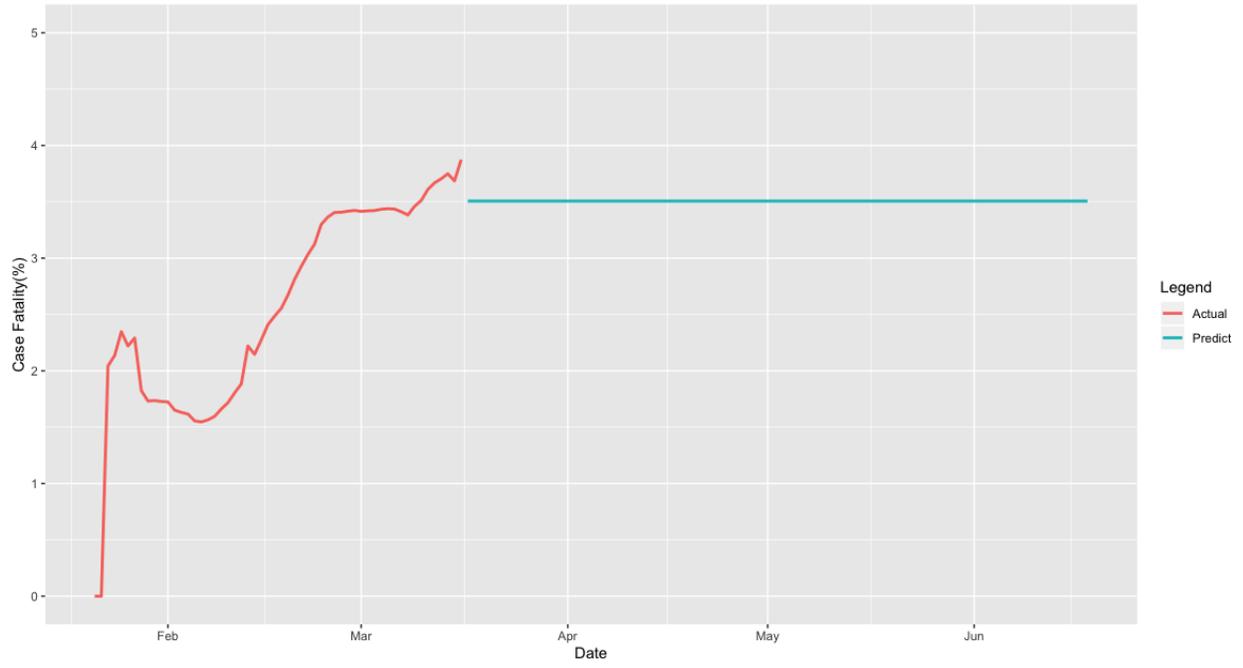

eTable 1. One- to five-step forecasting errors.

| | Reported | 1-step predicted | 1-step errors | 2-step error | 3-step error | 4-step error | 5-step error |
|---|---|---|---|---|---|---|---|
| **3/12/2020** | 125774 | 126272 | 0.40% | | | | |
| **3/13/2020** | 133774 | 130278 | -2.61% | 0.32% | | | |
| **3/14/2020** | 143864 | 144715 | 0.59% | -3.58% | -0.10% | | |
| **3/15/2020** | 155618 | 153628 | -1.28% | 2.51% | -4.08% | 0.03% | |
| **3/16/2020** | 170568 | 163932 | -3.89% | -2.02% | 2.26% | -4.80% | 0.34% |
| | **Average Absolute Error** | | 1.75% | 2.11% | 2.15% | 2.42% | 0.34% |

eTable 2. Spread of Covid-19 in 30 countries and world wide under two weeks delay intervention.

| State | Peak Time | End Time | Duration | Peak (Cum) | Peak (New) | Current Case | End Case |
|---|---|---|---|---|---|---|---|
| Total | 4/3/2020 | 9/11/2020 | 235 | 3657852 | 493023 | 170568 | 6522982 |
| ITALY | 4/3/2020 | 9/8/2020 | 222 | 727996 | 96948 | 24747 | 1307179 |
| SPAIN | 4/3/2020 | 8/20/2020 | 202 | 477245 | 64939 | 7753 | 852807 |
| IRAN | 4/3/2020 | 7/23/2020 | 155 | 413873 | 54653 | 14991 | 710755 |
| GERMANY | 4/3/2020 | 7/23/2020 | 178 | 309434 | 43000 | 4838 | 549478 |
| USA | 4/3/2020 | 9/2/2020 | 226 | 216943 | 29222 | 4740 | 381178 |
| FRANCE | 4/3/2020 | 8/18/2020 | 207 | 204820 | 27442 | 5380 | 363355 |
| SWIZTERLAND | 4/3/2020 | 9/6/2020 | 194 | 145504 | 20435 | 2200 | 257680 |
| BELGIUM | 4/3/2020 | 7/3/2020 | 150 | 132216 | 19337 | 1085 | 237907 |
| UK | 4/3/2020 | 9/11/2020 | 224 | 77356 | 10561 | 1395 | 138340 |
| NORWAY | 4/3/2020 | 6/22/2020 | 117 | 57561 | 8410 | 1077 | 105766 |
| AUSTRIA | 4/4/2020 | 7/23/2020 | 149 | 62095 | 7971 | 959 | 103793 |
| MALAYSIA | 4/4/2020 | 7/4/2020 | 161 | 57868 | 7303 | 553 | 93940 |
| CHINA | 2/5/2020 | 6/6/2020 | 138 | 31432 | 5236 | 81077 | 91305 |
| GREECE | 4/3/2020 | 7/20/2020 | 145 | 48448 | 6958 | 331 | 88863 |
| NETHERLANDS | 4/3/2020 | 8/2/2020 | 157 | 42387 | 5724 | 1135 | 75051 |
| PORTUGAL | 4/3/2020 | 6/22/2020 | 112 | 30116 | 4706 | 245 | 59791 |
| SLOVENIA | 4/3/2020 | 7/15/2020 | 133 | 27465 | 4190 | 219 | 56030 |
| ESTONIA | 4/3/2020 | 7/20/2020 | 145 | 27602 | 4221 | 205 | 55039 |
| FINLAND | 4/4/2020 | 7/22/2020 | 175 | 31047 | 4302 | 267 | 53472 |
| ISRAEL | 4/4/2020 | 6/30/2020 | 130 | 27678 | 3705 | 200 | 45801 |
| CZECHIA | 4/3/2020 | 6/20/2020 | 111 | 23470 | 3259 | 298 | 41366 |
| CANADA | 4/3/2020 | 7/8/2020 | 164 | 22704 | 3198 | 304 | 40782 |
| ICELAND | 4/4/2020 | 6/23/2020 | 114 | 22483 | 3087 | 138 | 37996 |
| BRAZIL | 4/3/2020 | 7/9/2020 | 134 | 17542 | 2509 | 200 | 34112 |
| ROMANIA | 4/4/2020 | 6/15/2020 | 110 | 19969 | 2759 | 158 | 33605 |
| QATAR | 4/3/2020 | 6/11/2020 | 103 | 18725 | 2701 | 401 | 33116 |

| | | | | | | |
|---|---|---|---|---|---|---|
| KOREA | 4/5/2020 | 5/31/2020 | 132 | 24100 | 1873 | 8236 | 31670 |
| AUSTRALIA | 4/3/2020 | 7/3/2020 | 161 | 16648 | 2310 | 298 | 29334 |
| POLAND | 4/3/2020 | 6/13/2020 | 102 | 13287 | 1908 | 150 | 24239 |
| INDONESIA | 4/4/2020 | 7/28/2020 | 149 | 12137 | 1811 | 117 | 23177 |

eTable 3. Spread of Covid-19 in top 30 countries and world wide under three weeks delay intervention.

| State | Peak Time | End Time | Duration | Peak (Cum) | Peak (New) | Current Case | End Case |
|---|---|---|---|---|---|---|---|
| Total | 4/10/2020 | 12/4/2020 | 319 | 16528763 | 2221889 | 170568 | 29313739 |
| ITALY | 4/10/2020 | 9/8/2020 | 222 | 3278431 | 439028 | 24747 | 5693059 |
| SPAIN | 4/10/2020 | 8/20/2020 | 202 | 2206610 | 297488 | 7753 | 3919623 |
| IRAN | 4/10/2020 | 8/21/2020 | 184 | 1882888 | 252756 | 14991 | 3360378 |
| GERMANY | 4/10/2020 | 8/31/2020 | 217 | 1426977 | 192760 | 4838 | 2521231 |
| USA | 4/10/2020 | 10/10/2020 | 264 | 992158 | 133015 | 4740 | 1801181 |
| FRANCE | 4/10/2020 | 9/23/2020 | 243 | 933029 | 124787 | 5380 | 1674855 |
| SWIZTERLAND | 4/10/2020 | 9/6/2020 | 194 | 677240 | 92147 | 2200 | 1210668 |
| BELGIUM | 4/10/2020 | 12/4/2020 | 304 | 620500 | 85031 | 1085 | 1114935 |
| UK | 4/10/2020 | 9/11/2020 | 224 | 354043 | 47385 | 1395 | 639280 |
| NORWAY | 4/10/2020 | 8/2/2020 | 158 | 266353 | 36481 | 1077 | 477043 |
| AUSTRIA | 4/10/2020 | 7/23/2020 | 149 | 251230 | 34028 | 959 | 446229 |
| MALAYSIA | 4/10/2020 | 8/7/2020 | 195 | 235384 | 32064 | 553 | 417237 |
| GREECE | 4/10/2020 | 8/23/2020 | 179 | 227442 | 30714 | 331 | 404882 |
| NETHERLANDS | 4/10/2020 | 11/8/2020 | 255 | 192602 | 25705 | 1135 | 353763 |
| PORTUGAL | 4/10/2020 | 7/5/2020 | 125 | 147451 | 20653 | 245 | 263904 |
| ESTONIA | 4/10/2020 | 8/4/2020 | 160 | 132648 | 18392 | 205 | 239417 |
| SLOVENIA | 4/10/2020 | 9/1/2020 | 181 | 130582 | 18060 | 219 | 236385 |
| FINLAND | 4/10/2020 | 8/30/2020 | 214 | 125404 | 17387 | 267 | 221710 |
| ISRAEL | 4/10/2020 | 8/13/2020 | 174 | 113848 | 15554 | 200 | 203392 |
| CZECHIA | 4/10/2020 | 7/7/2020 | 128 | 109112 | 14697 | 298 | 194123 |
| CANADA | 4/10/2020 | 10/17/2020 | 265 | 104602 | 14127 | 304 | 184096 |
| QATAR | 4/10/2020 | 8/19/2020 | 172 | 94166 | 13163 | 401 | 172432 |
| ICELAND | 4/10/2020 | 7/20/2020 | 141 | 94695 | 13174 | 138 | 167504 |
| ROMANIA | 4/10/2020 | 9/17/2020 | 204 | 82758 | 11451 | 158 | 147628 |
| BRAZIL | 4/10/2020 | 7/10/2020 | 135 | 81666 | 11128 | 200 | 145770 |
| AUSTRALIA | 4/10/2020 | 7/6/2020 | 164 | 77308 | 10457 | 298 | 138540 |
| CHINA | 2/5/2020 | 6/18/2020 | 150 | 31432 | 5236 | 81077 | 127241 |
| KOREA | 4/10/2020 | 7/29/2020 | 191 | 70196 | 8899 | 8236 | 123488 |
| POLAND | 4/10/2020 | 8/24/2020 | 174 | 62393 | 8447 | 150 | 111894 |
| EGYPT | 4/10/2020 | 8/10/2020 | 178 | 58876 | 8110 | 126 | 106174 |